# A solver for subsonic flow around airfoils based on physics-informed neural networks and mesh transformation

Wenbo Cao(曹文博), Jiahao Song(宋家豪), Weiwei Zhang(张伟伟)*

*School of Aeronautics, Northwestern Polytechnical University, Xi'an 710072, China.*

*Corresponding Author: Weiwei Zhang, aeroelastic@nwpu.edu.cn.*

**Abstract.** Physics-informed neural networks (PINNs) have recently become a new popular method for solving forward and inverse problems governed by partial differential equations (PDEs). However, in the flow around airfoils, the fluid is greatly accelerated near the leading edge, resulting in a local sharper transition, which is difficult to capture by PINNs. Therefore, PINNs are still rarely used to solve the flow around airfoils. In this study, we combine physical-informed neural networks with mesh transformation, using neural network to learn the flow in the uniform computational space instead of physical space. Mesh transformation avoids the network from capturing the local sharper transition and learning flow with internal boundary (wall boundary). We successfully solve inviscid flow and provide an open-source subsonic flow solver for arbitrary airfoils. Our results show that the solver exhibits higher-order attributes, achieving nearly an order of magnitude error reduction over second-order finite volume methods (FVM) on very sparse meshes. Limited by the learning ability and optimization difficulties of neural network, the accuracy of this solver will not improve significantly with mesh refinement. Nevertheless, it achieves comparable accuracy and efficiency to second-order FVM on fine meshes. Finally, we highlight the significant advantage of the solver in solving parametric problems, as it can efficiently obtain solutions in the continuous parameter space about the angle of attack.

## 1 Introduction

As scientific machine learning techniques, recently emerged methods for solving PDEs through neural network optimization, such as physics-informed neural networks (PINNs) [1], deep Ritz method [2], and deep Galerkin method [3], have been widely used to solve forward and inverse problems involving PDEs. By minimizing the loss function of PDE residuals, boundary conditions and initial conditions simultaneously, the solution can be straightforwardly obtained without mesh, spatial discretization, and complicated program. The concept of these methods can be traced back to 1990s, when neural algorithms for solving differential equations were proposed [4-8]. With the significant progress in deep learning and computation capability, a variety of PINN models have been proposed in the past few years, and have achieved remarkable results across a range of problems in computational science and engineering [9-12]. As a typical optimization-based PDE solver, PINNs offer a natural approach to solve PDE-constrained optimization problems. A promising application focuses on flow visualization technology [13-15], also referred to as data assimilation techniques [16-19], where flow fields or other unknown parameters can be easily inferred from sparse observations such as concentration fields and images. Such inverse problems are difficult for traditional PDE solvers. Moreover, several recent works [20-23] have demonstrated the significant advantages of PINNs in inverse design and optimal control, where solving the PDEs and obtaining optimal design are pursued simultaneously. In contrast, the most popular method, direct-adjoint-looping (DAL), involves hundreds or thousands of repeated evaluations of the forward PDE, yielding prohibitively expensive computational cost in solving PDE-constrained optimization problems. In addition, PINN-like methods have also achieved remarkable results in solving parametric problems [3, 24, 25].

Despite the potential for a wide range of physical phenomena and applications, training PINNs models still encounters challenges [9], and the current generation of PINNs is not as accurate or as efficient as traditional numerical method for solving many forward problems [11, 26]. There have been some efforts to improve PINNs trainability, which include adaptively balancing the weights of loss components during training [27-31], enforcing boundary conditions [8, 23, 32], transforming the PDE to its weak form [2, 33, 34], adaptively sampling [35, 36], spatial-temporal decomposition [37-39], using both automatic differentiation and numerical differentiation to calculate derivatives [40], and introducing pseudo-time derivatives to alleviate ill-conditioning [41].

Previous studies have observed that PINNs encounter challenges in convergence and accuracy when addressing problems with solutions containing sharp space transitions or fast time evolution, commonly referred to as "stiff problems" [34-36]. Such sharp transitions are more pronounced in the context of the flow around airfoils. To approximately simulate the flow around airfoils in an infinite region, we have to consider a computational domain that is large relative to the chord length of the airfoil, as shown in the Figure 1. In this flow, only the flow near the airfoil located in the center of the computational domain has a large flow gradient, especially at the leading edge. Such local sharp transition is difficult for PINNs to capture. In [25], the flow around airfoils is solved by considering only very small computational domain and extremely low Reynolds numbers, a setup that may lack practical significance in engineering. Despite the lack of reports, PINNs fail to solve the flow around such thin shape with far field boundary conditions, as we will see in Section 3.1 of this paper.

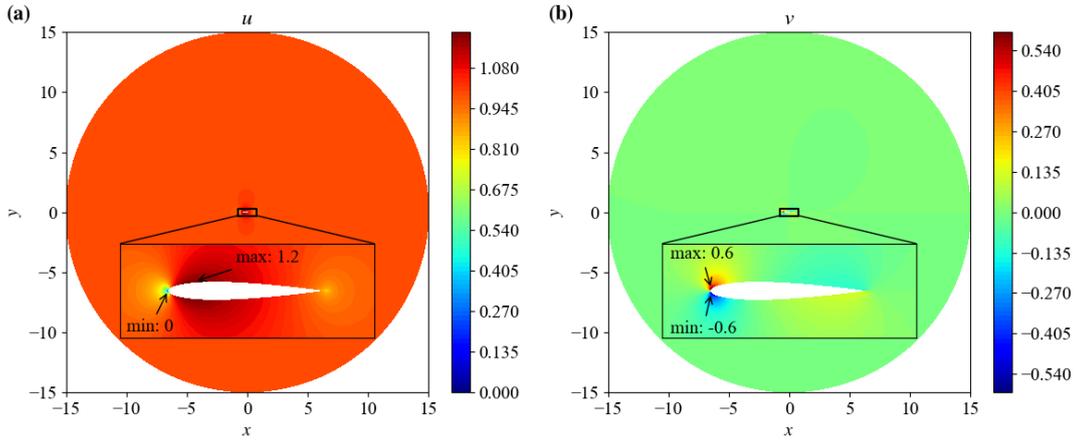

Figure 1. A typical flow around an airfoil, with a local enlargement of the flow near the airfoil inside the black outlined region. (a) The velocity component $u$. (b) The velocity component $v$.

In this study, we combine the physical-informed neural network with mesh transformation, using neural network learning the flow in the uniform computational space instead of physical space. The remainder of the paper is organized as follows. In Section 2, we begin by establishing the problem setting, followed by introducing mesh transformation and PINNs. Subsequently, we describe the challenges encountered by PINNs in solving airfoil flow in the physical space. Following this, we propose NNfoil, a PINN method combined with mesh transformation, to address these challenges. Section 3 enumerates extensive experimental studies to illustrate the advantages of NNfoil compared to PINNs and finite volume methods (FVM). Section 4 presents concluding remarks and directions for future research.

## 2 Methodology

### 2.1 Problem setting

We consider the two-dimensional Euler equation of inviscid flow, which is commonly used for rapid evaluation of aerodynamic force. A nonconservative, dimensionless typical form of Euler equation is

$$\frac{\partial Q}{\partial t} = -(A\frac{\partial Q}{\partial x} + B\frac{\partial Q}{\partial y}) \tag{1}$$

where $Q = [\rho \ u \ v \ p]^T$ is the vector of primitive variables; $\rho$ is the density; $u, v$ are the $x$-wise and $y$-wise components of the velocity vector $V$, respectively; $p$ is the pressure; and $A, B$ are the flux Jacobian, which have the following form

$$A = \begin{bmatrix} u & \rho & 0 & 0 \\ 0 & u & 0 & 1/\rho \\ 0 & 0 & u & 0 \\ 0 & \rho a^2 & 0 & u \end{bmatrix}, B = \begin{bmatrix} v & \rho & 0 & 0 \\ 0 & v & 0 & 0 \\ 0 & 0 & v & 1/\rho \\ 0 & 0 & \rho a^2 & v \end{bmatrix} \tag{2}$$

where $a^2 = \gamma p / \rho$ is the square of the sound speed; $\gamma = 1.4$ is the specific heat ratio. We consider the flow around airfoils. As shown in Figure 2, the flow approaches the uniform freestream conditions far away from the wall. Hence, the boundary conditions in the far field are

$$\rho_\infty = 1, u_\infty = \cos(\alpha), v_\infty = \sin(\alpha), p_\infty = 1/(\gamma Ma^2) \tag{3}$$

where $\alpha$ is the angle of attack and $Ma$ is the Mach number. Because the flow cannot penetrate the wall, the velocity vector must be tangent to the surface, and then the component of velocity normal to the surface must be zero. Let $\boldsymbol{n}$ be a unit vector normal to the surface as shown in Figure 2. The wall boundary condition can be written as

$$\boldsymbol{V} \cdot \boldsymbol{n} = 0. \tag{4}$$

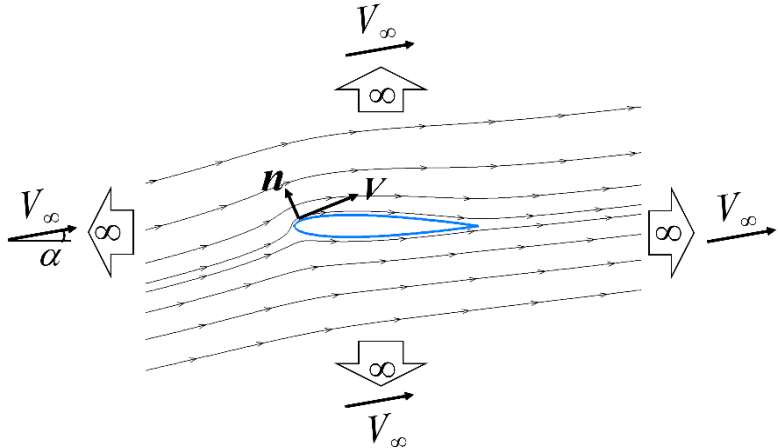

Figure 2. Boundary conditions at infinity and on the wall.

2.2 Mesh transformation

In the context of the finite difference method, there is no direct way to solve the governing equations over a nonuniform grid. Therefore, the non-uniform grid must be converted (somehow) into a uniform, rectangular grid. For the flow around an airfoil, a body-fitting coordinate system is usually used, as shown in Figure 3. $\Gamma_1$ is the wall boundary, $\Gamma_2$ is the far field boundary, and $\Gamma_3$ and $\Gamma_4$ coincide on the physical plane, satisfying periodic boundaries.

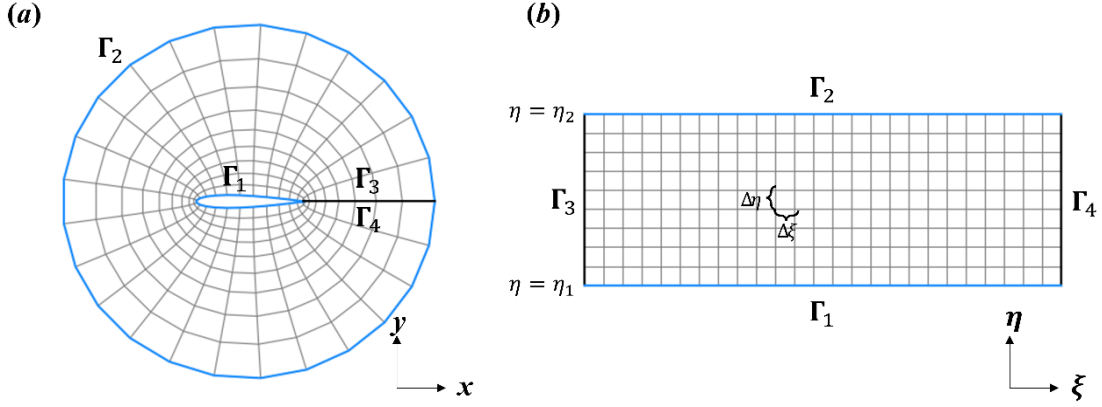

Figure 3. Schematic of a boundary-fitted coordinate system. (a) Physical space. (b) Computational space.

When the mesh transformation is applied, the derivatives in the physical space are evaluated by the transformation

$$\frac{\partial}{\partial x} = \frac{\partial}{\partial \xi}\frac{\partial \xi}{\partial x} + \frac{\partial}{\partial \eta}\frac{\partial \eta}{\partial x}$$
$$\frac{\partial}{\partial y} = \frac{\partial}{\partial \xi}\frac{\partial \xi}{\partial y} + \frac{\partial}{\partial \eta}\frac{\partial \eta}{\partial y} \tag{5}$$

where

$$\begin{bmatrix} \frac{\partial \xi}{\partial x} & \frac{\partial \eta}{\partial x} \\ \frac{\partial \xi}{\partial y} & \frac{\partial \eta}{\partial y} \end{bmatrix} = \begin{bmatrix} \frac{\partial x}{\partial \xi} & \frac{\partial y}{\partial \xi} \\ \frac{\partial x}{\partial \eta} & \frac{\partial y}{\partial \eta} \end{bmatrix}^{-1} = J^{-1} \tag{6}$$

The terms $\partial x/\partial \xi, \partial y/\partial \xi, \partial x/\partial \eta$ and $\partial y/\partial \eta$ can be evaluated by central differences. For more details to mesh transform, please refer to [42].

2.3 Physics-informed neural networks

In this section, we briefly introduce PINNs methodology. A typical PINN uses a fully connected deep neural network (DNN) architecture to represent the solution $q$ of the dynamical system. The network takes the spatial $x \in \Omega$ and temporal $t \in [0,T]$ as the input and outputs the approximate solution $\hat{q}(x,t;\boldsymbol{\theta})$. The spatial domain usually has 1-, 2- or 3-dimensions in most physical problems, and the temporal domain may not exist for steady (time-independent) problems. The accuracy of PINNs is determined by the network parameters $\boldsymbol{\theta}$, which are optimized with respect to PINNs loss function during the training process. To derive PINNs loss function, we consider $q$ to be mathematically described by differential equations of the general form:

$$\begin{aligned}&\mathcal{N}[q(x,t)] = 0, x \in \Omega, t \in (0,T]\\&\mathcal{I}[q(x,0)] = 0, x \in \Omega\\&\mathcal{B}[q(x,t)] = 0, x \in \partial\Omega, t \in (0,T]\end{aligned} \quad (7)$$

where $\mathcal{N}[\cdot]$, $\mathcal{I}[\cdot]$ and $\mathcal{B}[\cdot]$ are the PDE operator, the initial condition operator, and the boundary condition operator, respectively. Then PINNs loss function is defined as

$$\begin{aligned}&\mathcal{L} = \lambda_{PDE}\mathcal{L}_{PDE} + \lambda_{BC}\mathcal{L}_{BC} + \lambda_{IC}\mathcal{L}_{IC}\\&\mathcal{L}_{PDE} = \left\|\mathcal{N}[\hat{q}(\cdot;\boldsymbol{\theta})]\right\|_{\Omega\times(0,T]}^2\\&\mathcal{L}_{IC} = \left\|\mathcal{I}[\hat{q}(\cdot,0;\boldsymbol{\theta})]\right\|_{\Omega}^2\\&\mathcal{L}_{BC} = \left\|\mathcal{B}[\hat{q}(\cdot;\boldsymbol{\theta})]\right\|_{\partial\Omega\times(0,T]}^2\end{aligned} \quad (8)$$

The relative weights, $\lambda_{PDE}$, $\lambda_{BC}$, and $\lambda_{IC}$ in Equation (8), control the trade-off between different components in the loss function. The PDE loss is computed over a finite set of $m$ collocation points $D = \{x_i, t_i\}_{i=1}^{m}$ during training, along with boundary condition loss and initial condition loss. The gradients in the loss function are computed via automatic differentiation [43].

2.4 A PINN method combined with mesh transformation

As mentioned above, the leading-edge curvature of the airfoil is very large. The flow is greatly accelerated near the leading edge, yielding a local sharper transition, which is difficult to capture by PINNs. In addition, in inviscid flow, the signs of the velocity $v$ on the upper and lower surfaces of the airfoil are usually opposite, and the airfoil is very thin, which results in a weak discontinuity when the neural network learns the flow in the physical space. All these difficulties make PINNs fail to solve the flow around airfoils. When mesh transformation is performed based on a typical O-type grid (Figure 3), as shown in Figure 4, the local sharper transition in the physical space is smoothed in the computational space and dispersed to the entire domain, while avoiding the network learning the flow with an internal boundary (wall boundary condition).

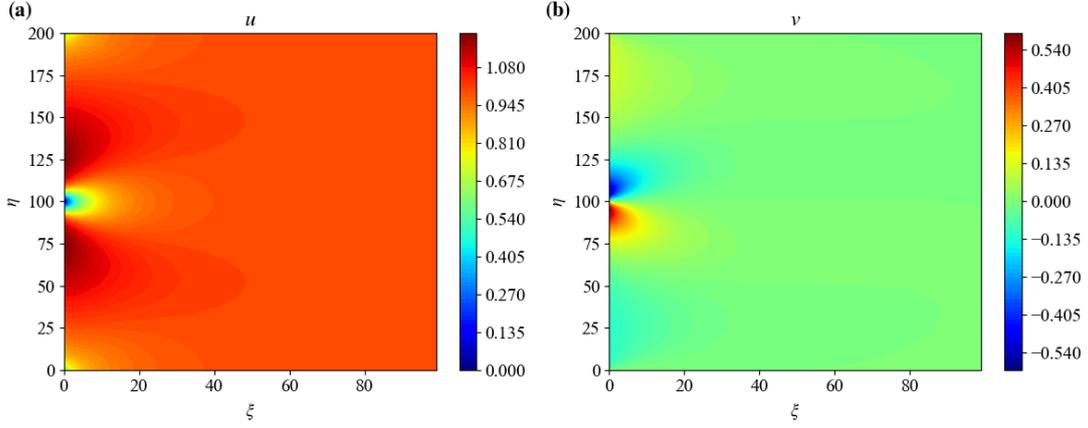

Figure 4. A typical flow around an airfoil in the computational space based on a typical O-type grid. (a) The velocity component $u$. (b) The velocity component $v$.

Therefore, we propose a PINN method combined with mesh transformation, referred to as NNfoil, meaning "neural network for airfoil". This method utilizes a neural network to learn the flow in the uniform computational space instead of physical space. As shown in Figure 5, the network takes $\xi, \eta$ as the input instead of $x, y$ and outputs the approximate solution at the grid points, and then evaluates the derivatives in the computational space using automatic differential (AD). Finally, calculate the derivatives and the loss in the physical space. In addition, we employ volume-weighted PDE residuals, which is shown to be better in non-uniform grids [44]. Thus, $\mathcal{L}_{PDE}$ is redefined as

$$\mathcal{L}_{PDE} = \|\mathcal{N}[\hat{q}(\cdot;\boldsymbol{\theta})]V\|^2 / \|V\|^2 \tag{9}$$

where $V$ is grid volume. We focus on the steady-state solution of the Euler equation, so the time derivative term in Equation (1) is omitted.

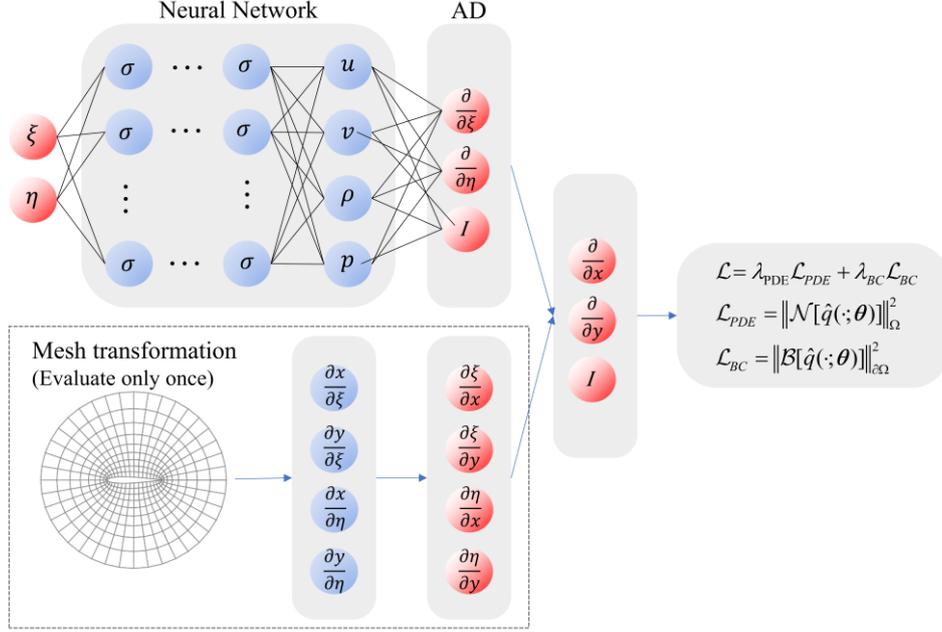

Figure 5. NNfoil, a PINN method combined with mesh transformation. Neural network is used to learn flow fields in the uniform computational space.

## 3 Results

In this section, we compare and discuss the efficiency and accuracy of NNfoil with the classical second-order finite volume method (FVM) under different flow conditions and airfoils. The FVM is one of the most widely used numerical methods in computational fluid dynamics (CFD), and it has extensive theoretical foundation for convergence. Thus, we use FVM on a very fine mesh ($N_\xi \times N_\eta = 800 \times 400$) to compute a very accurate reference solution as the ground truth. We solve the problem in a finite circular domain with a diameter $D_{far} = 30$, and the chord length of airfoils is 1. All our FVM calculations are performed on 4 Inter i9-13900KS CPU cores, while NNfoil calculations are performed on an NVIDIA GeForce RTX 4090 GPU.

Throughout all benchmarks, NNfoil will employ a fully connected DNN architecture with 5 hidden layers, each containing 128 neurons, equipped with the hyperbolic tangent activation functions (tanh), and trained using the limited-memory Broyden-Fletcher-Goldfarb-Shanno (LBFGS) optimizer. In PINN-like methods, the hyperparameters of the network and the relative weights between different losses significantly affect the convergence of training and the accuracy of solution. For the flow solver NNfoil, the hyperparameters can be selected according to the fact that the solution of a symmetric airfoil with zero angle of attack is symmetric. We choice $\lambda_{PDE} = 2 \times 10^4, \lambda_{BC} = 1$ in NNfoil. In all results, the FVM reduces the residuals by 10 orders of magnitude, while the NNfoil performs gradient descent for 20000 total iterations. The code and data accompanying this manuscript will be publicly available at https://github.com/Cao-WenBo/NNfoil.

3.1 A comparison of PINNs and NNfoil

In this subsection, we compare the results of PINNs and NNfoil in solving the flow. We consider shapes with different thicknesses, using the cylinder as the baseline and then gradually reducing its thickness (i.e., ellipses with different thicknesses). The flow conditions is $Ma = 0.4, \alpha = 0°$. The PDE loss are calculated on a mesh $N_\xi \times N_\eta = 200 \times 100$ for both PINNs and NNfoil. In the inviscid flow around an airfoil, the pressure coefficient distribution $C_p = (p - p_\infty)/(0.5\rho_\infty V_\infty^2)$ on the wall is the most concerned, which determines the lift coefficient $C_l$. Therefore, we evaluate the accuracy of different methods by relative $L_1$ errors of pressure coefficient and lift. The relative $L_1$ error between the predicted value $\hat{q}$ and the reference value $q_{ref}$ is defined as $\|\hat{q} - q_{ref}\|_1 / \|q_{ref}\|_1$.

As shown in Figure 6, we observe that for the flow around a cylinder (i.e., ellipse with thickness = 1), the results of both PINNs and NNfoil are consistent with the reference solution, and the error of the latter is lower. As the thickness decreases and the leading edge curvature increases, PINNs fail to capture the local sharper transition, while the result of NNfoil still consistent with the reference solution. For the ellipse with thickness = 0.4, we observe that the velocity of PINNs is asymmetric and the peak value is much smaller than the reference solution (Figure 7). Moreover, conventional airfoil thicknesses are typically less than 0.2 and have larger leading edge curvatures compared to the ellipse, which poses greater challenges for solving the flow using PINNs. Compared to PINNs, NNfoil still maintains the potential to solve the flow around thin airfoils, which will be discussed in more detail in the next section.

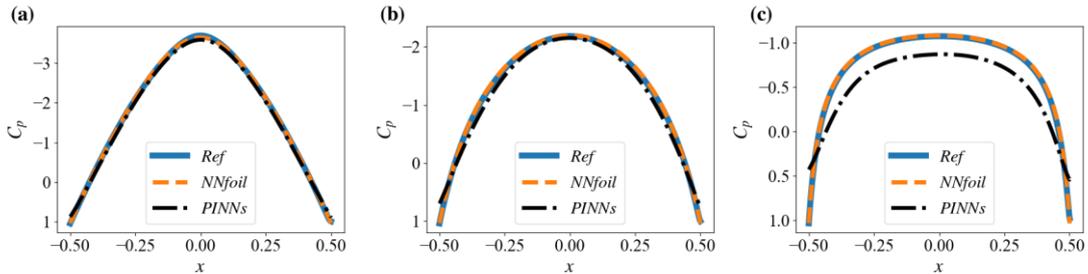

Figure 6. The pressure coefficient distributions of the flow around ellipses with thicknesses of (a) 1, (b) 0.7, (c) 0.4.

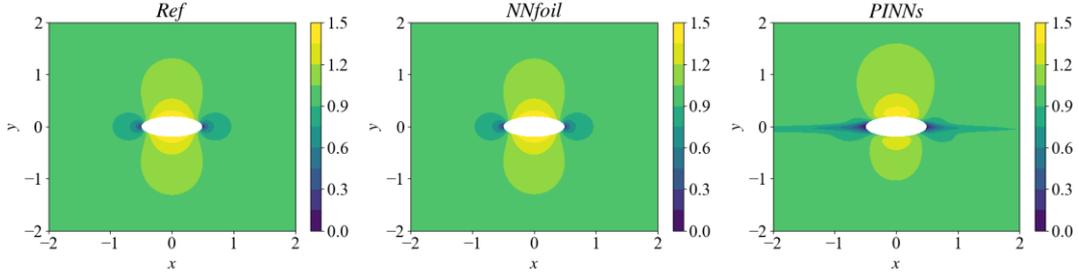

Figure 7. The flow fields of velocity $u$ around the ellipse with thickness = 0.4.

3.2 The flow around the NACA0012 airfoil

In this subsection, we consider the flow around the most classic NACA0012 airfoil with $Ma = 0.4, \alpha = 4°$, and study the convergence characteristics with mesh refinement. Figure 8 shows the pressure coefficients obtained by FVM and NNfoil with different meshes, respectively. For FVM, we observe that when the mesh is very sparse, the $C_p$ error is very large. As the mesh becomes denser, $C_p$ becomes more consistent with the reference solution, which is the typical convergence law of traditional grid-based numerical methods. For NNfoil, we observe that even when the mesh is extremely sparse, the error of the pressure coefficient distribution remains minimal, highlighting its high-order characteristics. However, as the mesh becomes denser, the error of NNfoil does not decrease gradually, unlike FVM. Although the network has no spatial discrete error and has higher accuracy with fewer meshes, its accuracy is also limited by the representation capabilities of neural networks and optimization algorithms. Despite this, the accuracy of NNfoil is satisfactory.

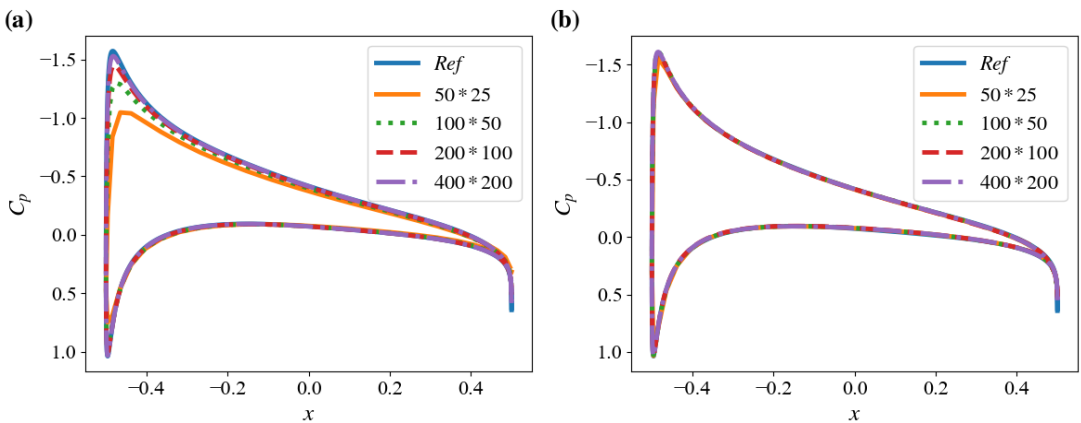

Figure 8. The pressure coefficient distributions of the flow around the NACA0012 airfoil obtained by (a) FVM, (b) NNfoil under different meshes.

The error and efficiency of FVM and NNfoil are compared in detail in Table 1.

Note that this table is not meant as a rigorous comparison of the computational efficiency of the two frameworks due to many different factors involved. In addition, we only compared traditional second-order FVM, which does not represent the latest advancements in high-order FVM, such as references [45, 46]. We observe that with the refinement of the mesh, the error of FVM decreases and the computing time increases. However, unlike FVM, for NNfoil, fewer meshes do not mean less computational costs. This is because the solution time of NNfoil is primarily influenced by the network architecture and optimization difficulty rather than the number of grids. Therefore, although NNfoil exhibits high-order characteristics, it may not be meant to be efficient, which is a departure compared to traditional numerical methods. In addition, we observed varying wall times for the same number of optimization steps, primarily because the LBFGS algorithm often stops early automatically.

Table 1: Error and computational time for obtaining the solutions of the flow around the NACA0012 airfoil by FVM and NNfoil.

| Mesh | FVM | | | NNfoil | | |
|---|---|---|---|---|---|---|
| | $C_p$ error | $C_L$ error | Wall time | $C_p$ error | $C_L$ error | Wall time |
| 50×25 | 0.176 | 0.184 | 9s | 0.029 | 0.022 | 384s |
| 100×50 | 0.077 | 0.038 | 24s | 0.015 | 0.010 | 266s |
| 200×100 | 0.033 | 0.012 | 53s | 0.015 | 0.018 | 262s |
| 400×200 | 0.011 | 0.004 | 285s | 0.014 | 0.011 | 271s |

3.3 Solutions for random flow conditions and airfoils

We next consider more general flow conditions and airfoils. Taking NACA0012 as the reference airfoil, 12 Class-Shape Transform (CST) parameters are used to parameterize the geometry, and 5 airfoils are randomly selected in the interval of ±30% disturbance, as shown in Figure 9. Then, 5 flow conditions are randomly selected in the interval $Ma \in [0.2, 0.7], \alpha \in [-10°, 10°]$ and combined with these airfoils as 5 cases, as shown in Table 2.

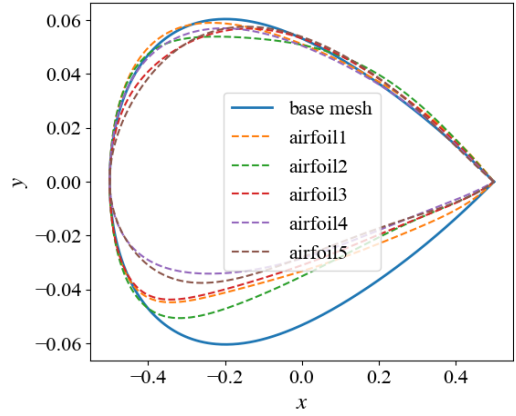

Figure 9. Airfoils randomly generated by perturbing CST parameters.

The results of FVM and NNfoil are shown in detail in Table 2. All examples use the mesh $N_\xi \times N_\eta = 200 \times 100$. We observe that except Case 4, the accuracy of NNfoil is better than that of FVM. Indeed, as verified in the previous subsection, the accuracy of NNfoil is roughly equivalent to the results of FVM on a very dense grid $N_\xi \times N_\eta = 400 \times 200$. These results validate that the results of NNfoil are reliable and satisfactory over a wide range of airfoils and flow conditions.

Table 2: Flow conditions and result errors obtained by FVM and NNfoil for different cases

| | Ma | $\alpha$ | Airfoil | FVM | | NNfoil | |
| --- | --- | --- | --- | --- | --- | --- | --- |
| | | | | $C_p$ error | $C_L$ error | $C_p$ error | $C_L$ error |
| Case1 | 0.28 | 6.1 | airfoil1 | 0.029 | 0.012 | 0.015 | 0.023 |
| Case2 | 0.24 | -3.0 | airfoil2 | 0.043 | 0.006 | 0.029 | 0.002 |
| Case3 | 0.41 | -3.5 | airfoil3 | 0.038 | 0.015 | 0.025 | 0.016 |
| Case4 | 0.68 | 1.8 | airfoil4 | 0.041 | 0.009 | 0.104 | 0.095 |
| Case5 | 0.34 | 4.9 | airfoil5 | 0.032 | 0.018 | 0.019 | 0.008 |

Figure 10 shows the pressure distribution. We observe that the $C_p$ of NNfoil at the suction peak (leading edge) is always closer to the reference solution compared to FVM, which once again verifies its high-order characteristics. In Case 4, we observed that the error of NNfoil is very large, which is because there is a shock wave in the flow, as shown in Figure 11, and the current version of NNfoil still fails to capture the shock wave. This difficulty may be solved in the future by using recent advances in the field such as [47, 48].

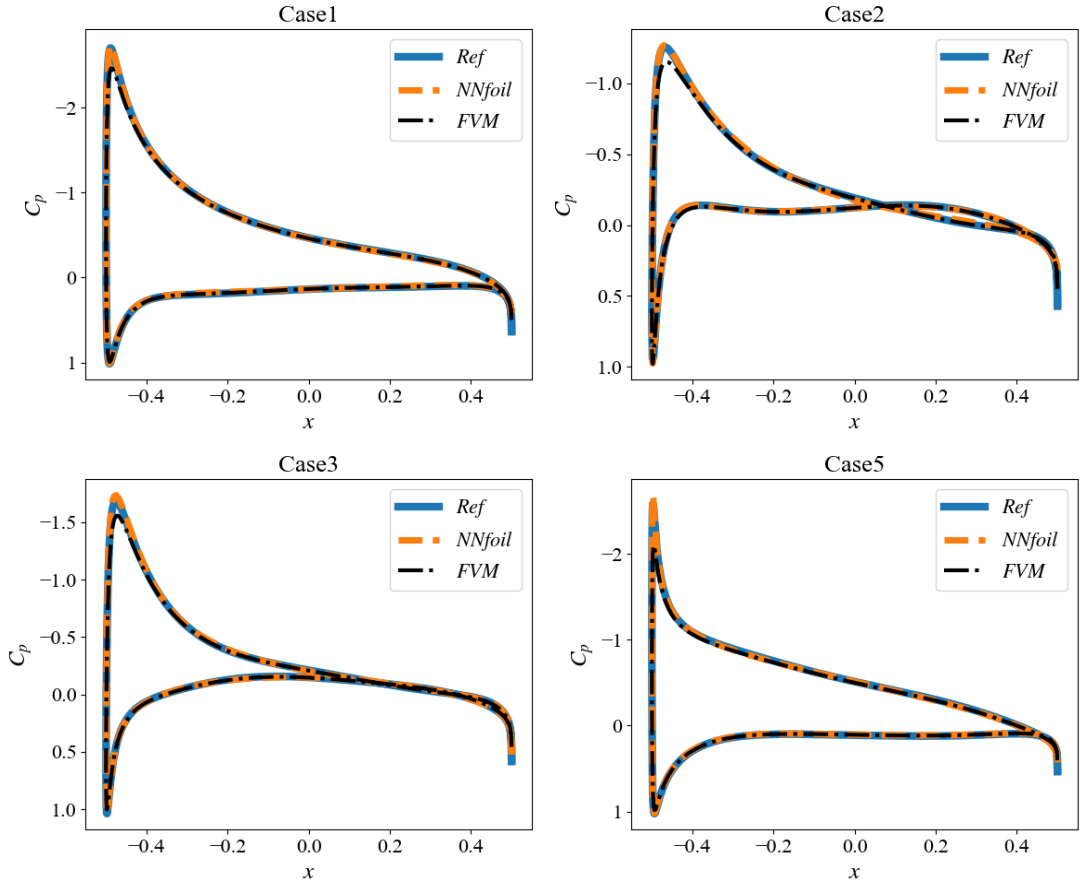

Figure 10. The pressure coefficient distributions obtained by FVM and NNfoil for different cases.

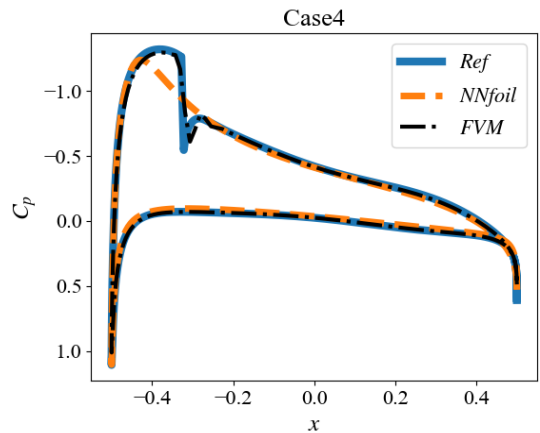

Figure 11. The pressure coefficient distributions obtained by FVM and NNfoil for Case 4.

3.4 Parametric problem about angle of attack

In aerodynamics, evaluating the performance of an airfoil often focuses on the pressure distribution and lift coefficient against angle of attack. In traditional CFD, the flow field of each angle of attack has to be solved separately, resulting in an expensive and repetitive task. However, in PINN-like methods, the more natural choice is to expand the neural network input dimension, leading to a parametric problem [20, 24]. In this study, we expand the input dimension of NNfoil from $[\xi,\eta]$ to $[\xi,\eta,\alpha]$, and then the collocation points are sampled in the joint parameter space of $[\xi,\eta,\alpha]$ to simultaneously solve the flow field with different angles of attack. Accordingly, the far field boundary conditions corresponding to each sample should be correctly processed, which are related to the angle of attack (Equation (3)). A significant advantage of this method is that it solves the flow in the continuous parameter space about angle of attack. It can be considered that the flow field at any angle of attack in the given interval is solved simultaneously, instead of only the flow at several discrete angles as in the traditional method.

We consider an asymmetric airfoil NACA2412, the flow condition is $Ma = 0.3, \alpha \in [-10°, 10°]$. To enhance the representation ability of the neural network for parametric problem, we increase its hidden layers from five to eight, and other settings are consistent with NNfoil. We perform gradient descent for 30000 total iterations using the LBFGS optimizer. Once the training is finished, the model can output the flow field at any angle of attack within the sampling interval. As shown in the Figure 12 and Figure 13, all results are consistent with the reference solution. More importantly, the model training wall time is 1543s, which is only about 5 times the time of solving the flow field at a single angle of attack time. However, for traditional numerical methods, to obtain the lift line, even if the solution interval for the angle of attack is 1 degree, the total calculation time is about 20 times the solution time for a single angle of attack.

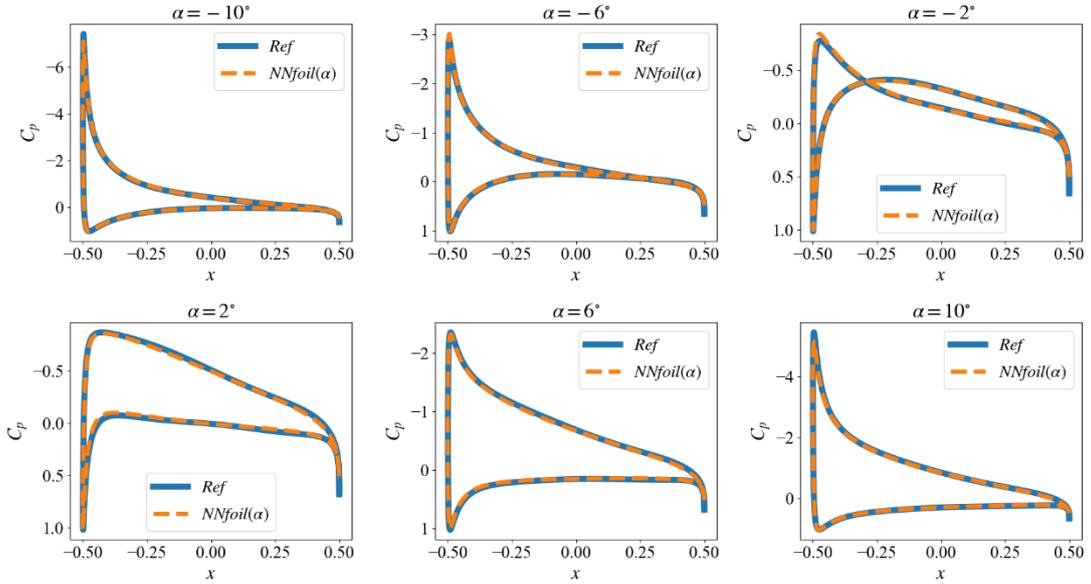

Figure 12. The pressure coefficient distributions of different angles of attack obtained parametric NNfoil, which is labeled as *NNfoil*($\alpha$) .

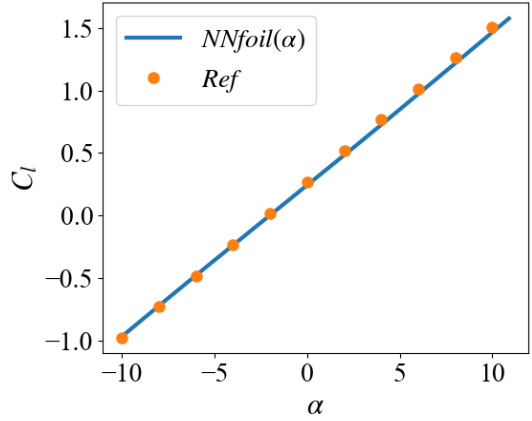

Figure 13. Lift line of parametric NNfoil. The reference solution is represented by discrete points, indicating that only specific angles of attack are solved using the FVM. In contrast, the result of parametric NNfoil is depicted by solid line, signifying its ability to derive a solution across the the continuous parameter space about the angle of attack. This implies that the parametric NNfoil can output the flow field and lift coefficient for any angle of attack within the solution interval.

## 4 Conclusions

In this work, we combine the physical-informed neural network the mesh transformation, using neural network to learn the flow in the uniform computational space instead of physical space. Mesh transformation avoids the network from capturing the local sharper transition and learning flow fields with internal boundary (wall boundary). We successfully solve the inviscid flow and provide a robust subsonic flow solver, capable of solving the flow around any airfoil with exactly the same settings, including network architecture, relative weights, and optimizer configuration. This solver achieves comparable accuracy and efficiency to second-order FVM on fine meshes, and can be used for rapid evaluation and analysis of aerodynamic forces of inviscid flow around airfoils. More importantly, as an optimization-based PDE solver, it has the potential to efficiently solve inverse problems related to the flow around airfoils, such as shape design optimization and optimal control. In addition, mesh transformation offers a viable approach for solving high Reynolds number flows in the future research. The most critical yet extremely thin boundary layer flows are amplified in the computational space by using appropriate mesh, avoiding the neural network to resolve very small-scale flows near wall in the physical space.

Although PINN-like methods are generally considered to lack advantages in forward problems, our results demonstrate several advantages of NNfoil over FVM in the flow around airfoils. The solver demonstrates higher-order characteristics on very coarse meshes, resulting in nearly an order of magnitude reduction in errors compared to second-order FVM. This improvement is attributed to the absence of discrete errors produced by automatic differentiation. Furthermore, the more important significance of NNfoil lies in its substantial advantage in solving parametric problems, as it can simultaneously and efficiently obtain solutions within the continuous parameter space of flow conditions, obviating the need to individually solve the flow for each flow conditions – a departure from conventional numerical methods.

Some problems remain open, e.g., limited by the learning ability and optimization difficulties of neural network, the accuracy of NNfoil will not increase significantly with mesh refinement. In addition, more complex flows remain to be addressed, such as compressible viscous flow, separated flow, supersonic flow, unsteady flow, and three-dimensional wing flow.

## Data Availability Statement

The data that support the findings of this study are available from the corresponding author upon reasonable request.

## Conflict of Interest Statement

The authors have no conflicts to disclose.

## Acknowledgments

We would like to acknowledge the support of the National Natural Science Foundation of China (No. 92152301).